\documentclass[%
 reprint,
superscriptaddress,
 amsmath,amssymb,
 aps,
prb,citeautoscript
]{revtex4-2}

\usepackage{graphicx}
\usepackage{dcolumn}
\usepackage{bm}
\usepackage{lipsum}
\usepackage{siunitx}
\usepackage{placeins}
\usepackage{amstext}
\usepackage{mathtools} 
\usepackage{amsmath,amssymb}    
\usepackage{booktabs}
\usepackage{array}
\usepackage{xcolor}
\usepackage{tabularx}
\usepackage{ulem}
\usepackage{hyperref}
\usepackage{nameref}

\usepackage{booktabs}

\DeclareSIUnit{\rad}{rad}
\DeclareSIUnit{\deg}{deg}

\hypersetup{
        colorlinks   = true,
        citecolor    = blue,
        linkcolor    = blue}

\DeclareSIUnit\bar{bar}
\DeclareSIUnit\torr{Torr}

\makeatletter
\def\@email#1#2{%
 \endgroup
 \patchcmd{\titleblock@produce}
  {\frontmatter@RRAPformat}
  {\frontmatter@RRAPformat{\produce@RRAP{*#1\href{mailto:#2}{#2}}}\frontmatter@RRAPformat}
  {}{}
}%

\let\svthefootnote\thefootnote
\newcommand\freefootnote[1]{%
  \let\thefootnote\relax%
  \footnotetext{#1}%
  \let\thefootnote\svthefootnote%
}

\begin{document}

\preprint{APS/123-QED}

\title{Cuprate Twistronics for Quantum Hardware}

\author{Tommaso Confalone}
\altaffiliation{\url{t.confalone@ifw-dresden.de}}
\affiliation{Leibniz Institute for Solid State and Materials Research Dresden (IFW Dresden), 01069 Dresden, Germany}
\affiliation{Institute of Applied Physics, Technische Universität Dresden, 01062 Dresden, Germany}

\author{Flavia Lo Sardo}
\affiliation{Leibniz Institute for Solid State and Materials Research Dresden (IFW Dresden), 01069 Dresden, Germany}
\affiliation{Institute of Materials Science, Technische Universität Dresden, 01062 Dresden, Germany}

\author{Yejin Lee}
\affiliation{Max Planck Institute for Chemical Physics of Solids, 01187 Dresden, Germany}

\author{Sanaz Shokri}
\affiliation{Max Planck Institute for Chemical Physics of Solids, 01187 Dresden, Germany}

\author{Giuseppe Serpico}
\affiliation{Department of Physics, University of Naples Federico II, 80125 Naples, Italy} 

\author{Alessandro Coppo}
\affiliation{Istituto dei Sistemi Complessi, Consiglio Nazionale delle Ricerche, 00185 Roma, Italy}
\affiliation{Department of Physics, University of Rome “La Sapienza”, 00185 Roma, Italy}

\author{Valerii\,M.\,Vinokur}
\affiliation{Terra Quantum AG, 9000 St.\,Gallen, Switzerland}

\author{Luca Chirolli}
\affiliation{Department of Physics and Astronomy, University of Florence, 50019 Sesto Fiorentino, Italy}

\author{Valentina Brosco}
\affiliation{Istituto dei Sistemi Complessi, Consiglio Nazionale delle Ricerche, 00185 Roma, Italy}
\affiliation{Department of Physics, University of Rome “La Sapienza”, 00185 Roma, Italy}

\author{Uri Vool}
\affiliation{Max Planck Institute for Chemical Physics of Solids, 01187 Dresden, Germany}

\author{Domenico Montemurro}
\affiliation{Department of Physics, University of Naples Federico II, 80125 Naples, Italy} 

\author{Francesco Tafuri}
\affiliation{Department of Physics, University of Naples Federico II, 80125 Naples, Italy} 

\author{Golam Haider}
\affiliation{Leibniz Institute for Solid State and Materials Research Dresden (IFW Dresden), 01069 Dresden, Germany}

\author{Kornelius Nielsch}
\affiliation{Leibniz Institute for Solid State and Materials Research Dresden (IFW Dresden), 01069 Dresden, Germany}
\affiliation{Institute of Applied Physics, Technische Universität Dresden, 01062 Dresden, Germany}
\affiliation{Institute of Materials Science, Technische Universität Dresden, 01062 Dresden, Germany}

\author{Nicola Poccia}
\altaffiliation{\url{nicola.poccia@unina.it}}
\affiliation{Leibniz Institute for Solid State and Materials Research Dresden (IFW Dresden), 01069 Dresden, Germany}
\affiliation{Department of Physics, University of Naples Federico II, 80125 Naples, Italy}


\begin{abstract}
Recent advances in the manipulation of complex oxide layers, particularly the fabrication of atomically thin cuprate superconducting films via molecular beam epitaxy, have revealed new ways in which nanoscale engineering can govern superconductivity and its interwoven electronic orders. In parallel, the creation of twisted cuprate heterostructures through cryogenic stacking techniques marks a pivotal step forward, exploiting cuprate superconductors to deepen our understanding of exotic quantum states and propel next-generation quantum technologies. This review explores over three decades of research in the emerging field of cuprate twistronics, examining both experimental breakthroughs and theoretical progress. It also highlights the 
methodologies poised to surmount the outstanding challenges in leveraging these complex quantum materials, underscoring their potential to expand the frontiers of quantum science and technology.
\end{abstract}

\maketitle

\section{Introduction}

\begin{figure*}
  \includegraphics[width=0.7\textwidth]{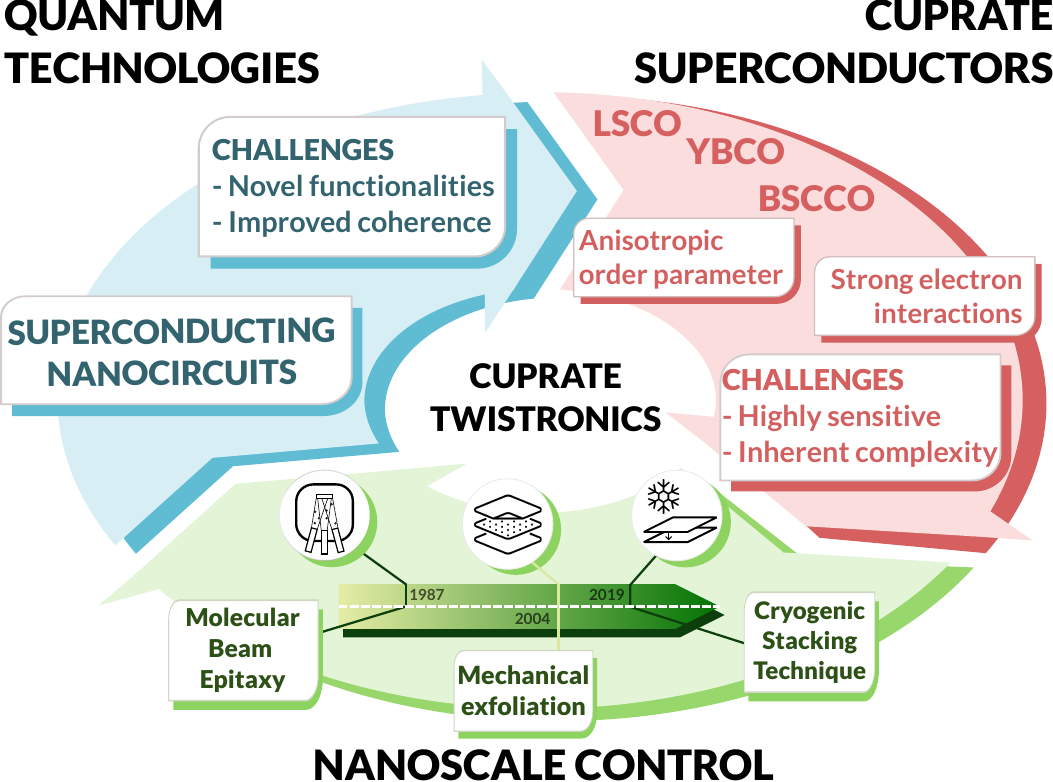}
  \caption{Overview of the three major fields intersecting in cuprate twistronics. Quantum hardware, particularly superconducting nanocircuits, faces the challenges of having novel functionalities and improved coherence, which may be addressed by integrating new materials like cuprates. High-temperature cuprate superconductors (BSCCO, YBCO, LSCO) exhibit inherent complexity and are highly sensitive, requiring advanced nanoscale control techniques. 
  Advances such as Molecular Beam Epitaxy, mechanical exfoliation, and cryogenic stacking have opened 
  possibilities for integrating these materials into quantum technologies.}
  \label{intro}
\end{figure*}

Quantum information science has witnessed remarkable progress over the past two decades, with multiple physical platforms emerging as candidates for scalable quantum technologies. Among them, superconducting qubits have gained significant traction due to their compatibility with existing semiconductor fabrication techniques, fast gate operations, and strong nonlinearities that enable high-fidelity control \cite{Kjaergaard2020, Krantz2019, Arute2019}. However, they typically operate at millikelvin temperatures and are limited by coherence times sensitive to material imperfections \cite{Gambetta2017, Muller2019}. In contrast, photonic systems excel in room-temperature quantum communication with minimal decoherence but face challenges in implementing deterministic interactions \cite{OBrien2009, Rudolph2017}. Trapped ions and neutral atoms offer exceptional coherence and well-controlled interactions yet are hindered by slower gate speeds and difficulties in large-scale integration \cite{Monroe2013, Saffman2010, Browaeys2020}. In this landscape, exploring high-temperature superconductors offers a promising avenue to bridge the advantages of superconducting circuits with enhanced operating temperatures and novel quantum phenomena. By pushing the boundaries of coherence and functionality at elevated temperatures, high-temperature superconducting platforms could dramatically reduce the cryogenic overhead, simplify system complexity, and open new pathways for hybrid quantum architectures.\\
Research into high-temperature superconductivity began with Bednorz and Müller’s 1986 discovery of superconductivity in cuprates, which still hold the record for the highest critical temperature ($T_c$) under ambient pressure \cite{Bednorz1986}. This discovery opened the door to a wide range of explorations across various materials, from magnesium diboride in 2001 \cite{Nagamatsu2001} to iron-based superconductors in 2006 \cite{Kamihara2006}, and, more recently, metal hydrides that exhibit superconductivity at near-room temperatures, albeit under extreme pressures \cite{Eremets2022}. While achieving higher $T_c$ values for practical applications has been a major focus, research is increasingly shifting toward the complex electronic behaviors of these materials, especially in copper oxides. Certain properties of cuprates, are now also observed in nickel-based materials, which exhibit similar crystal structures and electronic configurations but substitute nickel (Ni) for copper (Cu) \cite{Wang2024bis}. Recent studies on these nickelates have revealed unconventional superconducting characteristics, suggesting that they might share underlying mechanisms with cuprates, such as the d-wave order parameter \cite{Wang2024,Cheng2024}.\\
Cuprates, which are doped Mott insulators \cite{Lee2006}, still capture enormous interest because they exhibit unique properties that defy conventional theories \cite{Muller2014}, largely due to the strong electronic correlation that lead complex many-body quantum states. The Hubbard model is one theoretical framework used to partially explain this complexity, but it falls short of fully capturing the behavior of this “supreme” quantum matter \cite{Zaanen2024}. Its study connects with quantum information theory in multiple ways. Integrating unconventional superconductors into nanocircuits could enhance the coherence of superconducting qubits and unlock novel functionalities \cite{Tafuri2005, Tafuri2019, Ioffe1987, Brosco2024, Coppo2024, Patel2024}.  Additionally, understanding the quantum nature of the many-body wavefunction could provide 
insights into the fundamental physics and, at the same time, inspire strategies to harness the entanglement within these wavefunctions for quantum information applications \cite{Bellomia2024}. As a result, a wealth of research now prioritizes understanding and controlling the nanoscale electronic structures in cuprates. Thus, while raising $T_{c}$ remains an exciting avenue, the core intrigue lies in unraveling the deep, intricate physics of these superconductors, paving the way for 
quantum technologies that extend far beyond the pursuit of high-temperature thresholds alone.\\
Empirical observations indicate that materials discovered so far tend to achieve higher $T_{c}$ values as the number of elements in their chemical formulas increases. Based on experimental evidence, no binary compounds have been found to be superconducting above the boiling point of nitrogen at ambient pressure. This suggests that a certain inherent structural complexity, often described as optimal inhomogeneity, is required \cite{Poccia2012, Kivelson2007}. However, this complexity also limits our ability to precisely control these materials, which poses challenges for their use in applications like nanodevices. The competition between electronic phases such as superconductivity, magnetism, and charge order creates highly complex and unpredictable behaviors in materials with strong electron correlations \cite{Dagotto2005}. These materials often exhibit phase separation and nanoscale inhomogeneities, which pose significant challenges to theoretical modeling \cite{Carlson2015, Pelc2019}. In cuprates, for example, phase separation driven by factors like doping and internal chemical pressure significantly impacts superconducting transition temperatures \cite{Kugel2008}. Despite advancements in experimental techniques for exploring electron pairing, competing orders such as charge density waves \cite{Chaix2017} and unusual Fermi surface behaviors \cite{Meng2009}, a unified theory of high-temperature superconductivity remains elusive. Concepts such as quantum spin liquids, quantum criticalities and Lifshitz transitions, Fano resonances, minimal viscosity, multiband and quantum geometry are being investigated to deepen our understanding of superconductivity in these systems \cite{Keimer2015, Zhou2021, Bianconi2012}. However, the primary barrier to practical applications of cuprates may not be our lack of theoretical understanding. Instead, the real challenge lies in the difficulty of controlling and maintaining the specific nanoscale conditions required for their use in advanced quantum technologies.\\
The observation of atomically thin superconductivity in cuprates, achieved through molecular beam epitaxy (MBE), marks a key step toward control of superconductivity in nanoscale applications \cite{Webb1987, Tsukada1996}.
Using MBE it was possible to make thin films in an atomic layer of La$_{2-x}$Sr$_{x}$CuO$_{4}$ (LSCO), a cuprate material, demonstrating precise control over the thickness and composition of the films, isolating superconductivity from a single plane \cite{Bollinger2016}. It revealed how the spatial arrangement of atoms and interface effects influence the superconducting phase, offering 
insights into controlling these exotic electronic states by material engineering \cite{Rimal2024}. Although LSCO at a doping level of $x \approx 0.15$ achieves a superconducting transition temperature of around 40\,K, this is still quite below the nitrogen boiling point (77\,K). This contrasts with other high-$T_{c}$ cuprates, such as Bi$_{2}$Sr$_{2}$CaCu$_{2}$O$_{8+x}$ (BSCCO), which exhibit higher $T_{c}$ values, sometimes approaching 90\,K or more. However, the fabrication of atomically thin layers of BSCCO through techniques such as MBE and pulsed laser deposition (PLD) faces significant challenges \cite{Yelpo2020}.\\
The complex chemistry of BSCCO—specifically its layered structure and the need for precise stoichiometric control—makes the realization of atomically thin systems more difficult compared to other cuprates like LSCO. In the early 2000s, the isolation of graphene using a scotch tape by Andre Geim and Konstantin Novoselov marked a revolution in material science, transforming our understanding of how to create atomically thin materials, termed two-dimensional (2D) materials \cite{Novoselov2005}. Their groundbreaking work on graphene's exceptional conductivity, which awarded them the Nobel Prize in 2010, inspired further exploration into other 2D materials. One such study focused on BSCCO, revealing its insulating properties and large band gaps, which at that time set it apart from materials like graphene, which exhibited metallic conductivity \cite{Novoselov2005}. It was noted that the scotch tape could simply break the van der Waals bond of the BiO layers and stamp on a substrate many BSCCO crystals of different thicknesses. However, a first understanding that not all 2D materials stamped on a substrate such as graphene are inert and stable in the atomically thin limit came from micro-Raman experiments performed on exfoliated BSCCO crystals \cite{Sandilands2010}, highlighting factors affecting their stability such as oxygen out-diffusion and environmental degradation. The study noted that aging and exposure to air could lead to degradation, affecting the Raman features and indicating a transition toward an amorphous structure in some samples.\\
An initial recognition that exfoliated flakes of BSCCO crystals could be used in quantum technologies, arose with experimental work showing that high temperature superconductivity was likely induced in Bi$_{2}$Se$_{3}$ and Bi$_{2}$Te$_{3}$ through proximity to BSCCO \cite{Zareapour2012}. This discovery underscored the necessity of extremely clean interfaces, driving further exploration of BSCCO cuprate interfaces and the technologies required to control their quantum complexity. However, the environmental sensitivity of atomically thin BSCCO layers and other 2D materials beyond graphene presented a challenge, prompting researchers to develop methods for device fabrication while preventing material degradation from exposure to air and moisture. One effective solution was the use of inert gas-filled glovebox systems, which provide an inert atmosphere for handling sensitive materials \cite{Gray2020}. This approach eventually led to the development of cleanroom-in-a-glovebox setups, which integrate the controlled environment of a glovebox with the advanced tools of a cleanroom, allowing precise manipulation and fabrication of 2D materials without compromising their integrity.\\
Between 2018 and 2019, pioneering experiments were conducted to measure the electrical properties of exfoliated thin BSCCO crystals within a glovebox atmosphere. These studies explored different exfoliation techniques and circuit designs, which are key to our fundamental understanding of these materials \cite{Liao2018,Yu2019,Zhao2019}. In one method, researchers used prepatterned contacts to create a four-unit cell crystal with superconducting properties. Using a solid-ion conductor substrate, they could switch the crystal between superconducting and insulating states, demonstrating the potential for electrostatic gating, though lithium ions sometimes penetrated and altered the crystals \cite{Liao2018}. A second approach produced a Hall-bar nanodevice by exfoliating crystals onto a silicon oxide substrate. Even in crystals as thin as two-unit cells, superconducting properties were preserved, with fluctuations and vortex effects observed to become more dominant. Silicon nitride stencil masks, ultrahigh-vacuum contact evaporation, and boron nitride encapsulation helped maintain material integrity \cite{Zhao2019}. In a third approach, a 0.5-unit cell cuprate crystal was exfoliated at -40$\,^\circ$C, preserving bulk-like superconducting properties. Here, cold-welded indium wire formed the macroscopic device bonds, and gating was achieved by carefully adjusting the oxygen content in an ozone atmosphere, offering a unique method to control intrinsic properties compared to other 2D metallic systems \cite{Yu2019}. Scanning tunneling microscopy (STM) experiments confirmed that typical bulk electronic patterns and inhomogeneities were retained, underscoring the remarkable superconductivity of BSCCO at atomic thickness due to its extremely short coherence length.\\
In a completely original 
approach, the previous three methods for synthesizing atomically thin crystals were combined to stack and twist thin BSCCO crystals into heterostructures. This technique, termed Cryogenic Stacking Technology (CST), was first introduced on arXiv in 2021 \cite{Zhao2021} and later detailed in three separate publications in 2023 \cite{Zhao2023,Martini2023,Lee2023}. CST enables the creation of ultraclean junctions at various twist angles under ultralow temperatures (below -90$\,^\circ$C), in an inert atmosphere, using stencil mask evaporation in ultra-high or high vacuum with boron nitride encapsulation. This method represents a significant fresh 
direction over MBE or PLD for nanoscale manipulation of cuprate superconductors. This breakthrough lays the groundwork for the continued pursuit of utilizing cuprate materials in active quantum hardware, a goal that has long been sought and is the main focus of this review. Figure (\ref{intro}) provides a comprehensive overview of how the three key fields discussed so far intersect, emphasizing the challenges within each field that can be addressed by exploring solutions from the others.\\
This review is organized into three main sections and a concluding outlook. In the first section, we examine the past 30 years of experimental research on intrinsic Josephson junctions in cuprates and twisted cuprate heterostructures along the vertical c-axis. The second section focuses on the theoretical work that provides a comprehensive framework for understanding twisted cuprate junctions and their potential to reveal 
unique fundamental phenomena. In the third section, we discuss both experimental and theoretical efforts aimed at developing quantum technologies using high temperature cuprate superconductors. Finally, the outlook briefly addresses the current challenges for applying cuprate twistronics in advanced quantum hardware.\\
\begin{figure*}
  \includegraphics[width=0.9\textwidth]{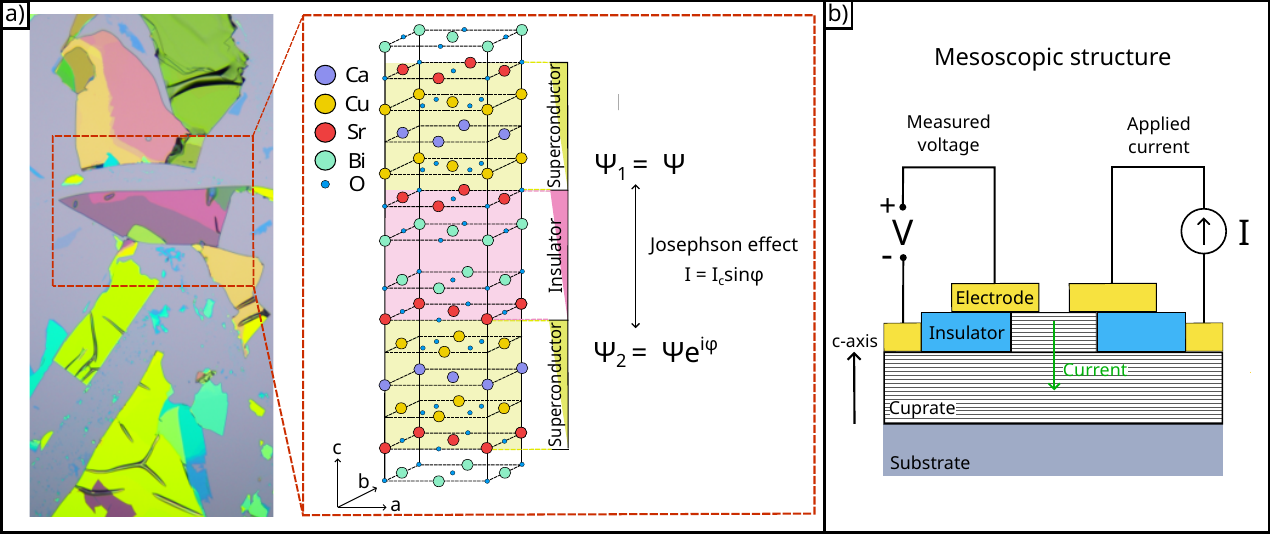}
  \caption{a) Optical image of mechanically exfoliated Bi$_{2}$Sr$_{2}$CaCu$_{2}$O$_{8+x}$ (BSCCO) flakes on a standard SiO$_2$/Si substrate. The color variations correspond to differences in thickness, ranging from $\sim$10\,nm to $\sim$200\,nm. The inset highlights the material's crystal structure, which is representative of the cuprates structure. The layered arrangement is visible, resembling a stack of superconducting/insulating layers, thus forming intrinsic Josephson junctions ($I_c$ is the critical current and $\varphi$ is the phase difference). b) A schematic of typical mesa structure used for probing the electronic properties of the intrinsic stack in cuprates.}
  \label{IJJs}
\end{figure*}

\section{Experimental advances in Cuprate twistronics}

\subsection{The subtlety of temperature degradation} 
Josephson junctions (JJs) are at the heart of modern superconducting circuits, playing a crucial role in quantum computing architectures, ultra-sensitive magnetic sensors, and high-frequency electronics \cite{Citro2024}. These advancements are based on tunable nonlinear quantum mechanical characteristics of JJs. Specifically, a JJ is a quantum mechanical system formed by two superconductors separated by a thin barrier or a weak link, typically an insulator or a non-superconducting metal. It allows the tunneling of Cooper pairs through the barrier without resistance. The behavior of the junction is described by the Josephson effect, predicted by Brian D. Josephson in 1962 \cite{Josephson1962}, the main feature of which is the ability to sustain a current up to a maximum value called critical current ($I_c$) without developing a voltage.\\ 
Since their discovery, cuprate superconductors have been recognized as exceptional due to their high critical temperatures and other properties that challenge conventional theories \cite{Muller2014, Habermeier2016}. Notably, these materials naturally form JJs within their crystal structures. In 1992 was demonstrated that the layered structure of Bi$_{2}$Sr$_{2}$CaCu$_{2}$O$_{8+x}$ (BSCCO) facilitates intrinsic Josephson junctions (IJJs): each CuO$_{2}$ layer acts as a superconductor, while the intervening rocksalt layer serves as an insulating barrier, creating a stack of JJs along the c-axis without the need for additional fabrication \cite{Kleiner1992}. By applying an electric current along this axis, they observed Josephson-like voltage steps—evidence of coherent Cooper pair tunneling between the layers. Subsequently, the intrinsic Josephson effect was also confirmed in YBa$_{2}$Cu$_{3}$O$_{7-y}$ (YBCO) \cite{Rapp1996} and La$_{2-x}$Sr$_{x}$CuO$_{4}$ (LSCO) \cite{Uematsu1998}, establishing intrinsic tunneling as a defining feature of cuprate superconductors. BSCCO remains a preferred choice for exploring IJJs due to its ease of cleaving along the c-axis. Figure (\ref{IJJs}) depicts the structure of Bi$_{2}$Sr$_{2}$CaCu$_{2}$O$_{8+x}$ (BSCCO) as an example. It highlights how the layered configuration creates a stack of intrinsic Josephson junctions and demonstrates how the electrical properties of this stack can be investigated through transport measurements.\\
This discovery triggered a series of experiments on mesa structure fabrication of IJJs, grouped in a later review \cite{Yurgens2000}, with the aim of studying high-temperature superconductivity in cuprates. In the beginning, experiments were focusing on whether the superconducting order parameter (SOP) in BSCCO includes an isotropic s-wave or d-wave component by realizing twisted cuprate junctions. The main idea was to see whether, by twisting the layers with respect to each other, the critical current density would vary with the twist angle or not, distinguishing between the two order parameters. These studies can be seen as the c-axis counterpart to the experiments conducted in the ab-plane with grain boundary junctions, discussed in Section \ref{sec:GBJJs} of this review.\\
In a first experiment, the fabrication was performed using high-quality single crystals of BSCCO, prepared with the traveling solvent floating zone technique \cite{Li1999}. These crystals were then cleaved in the ab-plane in air to create smooth surfaces. One cleaved crystal was rotated by a desired angle around the c-axis relative to another identical cleaved crystal. The two cleaved crystals were stacked with their twist alignment and subjected to a controlled sintering process: a junction between two thick bulk crystals was then formed. This involved heating the stacked crystals in controlled oxygen pressure for 30 hours at a temperature just below the melting point of BSCCO (approximately 885$\,^\circ$C to 890$\,^\circ$C, depending on stoichiometry). The electrical contacts were then realized with silver paste and epoxy. The resulting measurements showed that the critical current density was unaffected by the twisting angle, supporting the hypothesis of either purely incoherent tunneling across layers or the presence of an isotropic order parameter.\\
The same experiment was performed with whiskers of a few microns thickness instead of single crystals \cite{Takano2002}. It was pointed out that, for creating the junctions, annealing at high temperature for excessive time would result in a welding rather than a joining of the BSCCO crystals. In this case, BSCCO whiskers were grown from an amorphous BSCCO precursor plate prepared via a melt-quench process and then annealed at 850$\,^\circ$C for 120 hours in an atmosphere of $72\%$ oxygen and $28\%$ nitrogen. However, these crossed whiskers were then annealed at 850$\,^\circ$C for a much shorter time, around 30 minutes, in a $70\%$ oxygen and $30\%$ argon atmosphere to join them. This short annealing process ensured effective contact without overheating, allowing the whiskers to form a clean and stable junction. Gold leads were attached to the ends of each whisker with silver paste for electrical measurements. The result showed a critical current density dependence consistent with an anisotropic SOP hypothesis. However, the angular dependence was much stronger than that of the conventional $d$-wave, as recognized by the same authors.\\
A follow-up experiment was subsequently conducted to investigate c-axis transport in BSCCO cross-whisker junctions naturally grown \cite{Latyshev2004}. During the growth process at temperatures close to the melting point of the BSCCO, some whiskers naturally intersected, forming cross-junctions with contact between their ab-faces. Using naturally grown cross-junctions would eliminate unwanted effects from the substrate on which they were previously deposited. After their formation, an annealing process was performed around 845$\,^\circ$C for 20 minutes to reduce the high interface resistance that these cross-junctions exhibited. In this case, the critical current density was found to be independent of the twist angle, enforcing instead the conclusion of a dominant $s$-wave component.\\
In the following years, it became evident that BSCCO was highly chemically sensitive when reduced to a thin crystal. As a result, all experiments involving the assembly of twisted cuprate heterostructures were conducted in a glovebox environment to minimize moisture exposure. Additionally, it became clear that thinner BSCCO crystals were increasingly prone to chemical and structural instability, making them more susceptible to degradation. Flakes obtained through mechanical exfoliation using the scotch tape method typically measure a few tens of nanometers in thickness, further increasing their vulnerability to environmental degradation.\\
In one attempt to explore the twistronics of BSCCO flakes, bulk single crystals of nearly optimally doped BSCCO were grown by the traveling floating zone method \cite{Zhu2021}. They were subsequently mechanically exfoliated to obtain ultrathin flakes in a glovebox with an Ar atmosphere. Two designated BSCCO flakes were sequentially placed on top of a SiO$_2$/Si substrate with prepatterned electrodes, realizing in this way the twist junction. The samples were then annealed for approximately 10 minutes, either in flowing oxygen at ambient pressure at 530$\,^\circ$C or in ozone at a pressure of around 5e$^{-5}$\,mbar at 450$\,^\circ$C. Double steps transitions in the superconducting characteristics are observed, indicating possible degradation. The results indicate that no angular dependence is observed as a function of the twist angle, suggesting a dominant $s$-wave SOP.\\
In a following attempt, it was recognized that the fabrication of an atomically clean and sharp junction interface is crucial for investigating the angular dependence of the SOP in BSCCO \cite{Lee2021}. Here a polymer adhesive stamp (thermoplastic methacrylate copolymer, Elvacite 2552C) was used to cleave a single flake into two pieces at a controlled temperature ($\sim$100$\,^\circ$C), forming then the junction. The stacking process was done in an argon-filled glovebox within 10 minutes. However, in this experiment, residual polymer from the adhesive stamp was removed using acetone and annealing was performed at 350$\,^\circ$C for 30 minutes in an oxygen atmosphere. Moreover, the Ag/Au electrodes were evaporated and annealed again at 350$\,^\circ$C  to diffuse Ag into the BSCCO for better electrical contact. These additional steps could have introduced additional chemical and temperature degradation. On the other hand, the lower temperature and timing in the fabrication of the interface likely improved the cleanliness of the interface as an anisotropic angular dependence, very similar to the one of the whisker junctions, was observed.\\
From this series of experiments on twisted cuprate heterostructures, some preliminary observations can be made. First, in all works, the interface was not superconducting after the fabrication process, and thus an annealing process, long or short, was needed to restore superconductivity. Secondly, in all the experiments discussed, the electrical contacts relied on prepatterned pads or on some heavy use of solvents and polymers.\\

\begin{figure*}
  \includegraphics[width=0.8\textwidth]{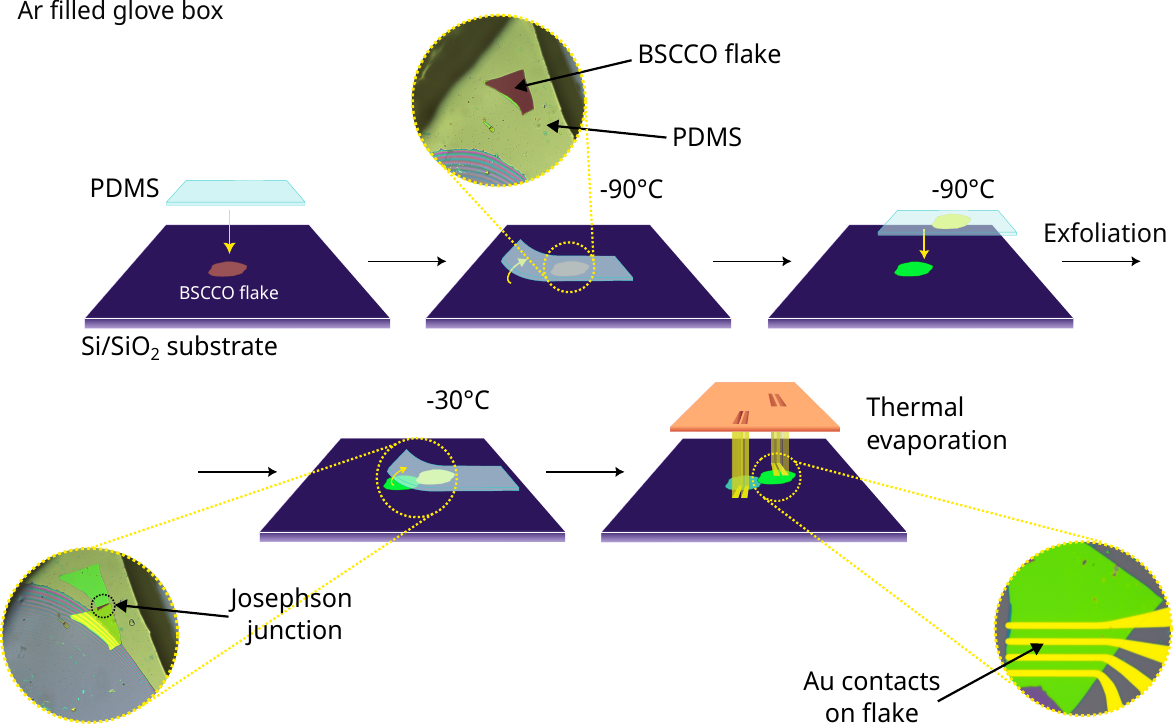}
  \caption{Schematic of the cryogenic stacking technique (CST) for fabricating c-axis BSCCO Josephson junctions (JJs). A Polydimethylsiloxane (PDMS) stamp is positioned on a mechanically exfoliated BSCCO flake. At -90$\,^\circ$C, the PDMS is dragged to separate the flake into two parts: a "bottom" flake that remains on the substrate and a "top" flake that adheres to the stamp. The top flake is then quickly aligned and placed back onto the bottom flake to form the JJ. Gradually increasing the temperature to -30$\,^\circ$C allows the PDMS stamp to be removed from the stacked flakes. Finally, thermal evaporation, combined with a stencil mask technique, is used to create electrical contacts on the JJ.}
  \label{fabrication}
\end{figure*}

\subsection{The crucial role of interstitial oxygen order in cuprate}
Recognition that generally BSCCO crystals should not be exposed to solvents and chemicals during the fabrication of Josephson junctions arrived from experiments where X-ray nanobeams were used as patterning tools \cite{Pagliero2014, Truccato2016}. Here the experimental setup utilized Kirkpatrick-Baez mirror optics in pink-beam mode at a synchrotron radiation facility, achieving beam sizes of 50\,nm$^2$ at about 17\,keV. Using this 
method, JJs were created, but resistance increased by approximately $50\%$  across all temperatures after irradiation, indicating modified electrical properties. The superconducting transition also shifted, exhibiting a long tail with nonzero resistivity down to 70\,K, suggesting structural changes, likely due to oxygen loss. However, it was already well understood that oxygen loss in a cuprate superconductor is just one of the most apparent forms of lattice degradation. Oxygen ordering and disordering are subtle forms of change in the oxygen-defect self-organization of cuprate crystals that can also have a profound effect on the general superconducting properties of the system. It might, therefore, not be surprising that a careful control not only of the oxygen loss, but also of the oxygen ordering might be important to consider in achieving structurally coherent and cleaner twisted cuprate Josephson junctions.\\
In some cuprates, such as the La$_2$CuO$_{4+y}$, the material is closer to a structural transition between tetragonal and orthorhombic phases than BSCCO. This transition allows greater mobility of interstitial oxygen dopants, and therefore a macroscopic phase separation as evidenced in multiple experiments \cite{Statt1995, Jorgensen1987,Beyers1989}, making the physics of the defects in cuprate more easily visible and less subtle. In La$_2$CuO$_{4+y}$, for example, the ordering of the interstitial oxygen dopants is specifically influenced by the tilt of the CuO$_{6}$ octahedra. A uniform distribution of tilt angles, modulated linearly along a crystal direction, creates structural openings in the LaO layers. This facilitates oxygen ordering into stripe-like patterns.\\
This special order was studied using X-ray microdiffraction ($\mu$XRD). Single-crystal samples of La$_2$CuO$_{4+y}$ were cleaved along the CuO$_{2}$ planes, and the integrated satellite peak, linked to the three-dimensional Q2 superstructures of the ordered interstitial oxygen dopants, was mapped. The ordering was reflected by the variation in the intensity of the Q2 superstructure and quantified by a power-law distribution, indicating a fractal nature in real space. This fractal nature is also supported by the power law dependency of the spatial correlation function \cite{Fratini2010}. Additionally, the study examines how thermal treatments affect the ordering of interstitial oxygen (i-O), with annealing conditions ranging from no higher than 50$\,^\circ$C to no lower than -73$\,^\circ$C. Different superconducting phases and Q2 intensity behaviors were observed depending on the specific conditions \cite{Fratini2010}. Similar results were noted for the oxygen interstitial 1D chains in Hg$_2$CuO$_{4+y}$ \cite{Campi2015}. The material displays characteristics fractal orders, disordered or ordered at similar temperatures than in La$_2$CuO$_{4+y}$, strongly influencing the $T_c$ of the samples despite unchanged doping level. Both works demonstrate that the oxygen interstitials ordering is highly inhomogeneous at the micrometer scale while simultaneously exhibiting a fractal nature, beneficial to the superconducting properties of the cuprates \cite{Zaanen2010, Yuan2024}. This inhomogeneity is also clearly observed in underdoped YBa$_2$Cu$_3$O$_{6+x}$ samples, where high-energy X-ray diffraction has revealed four distinct Cu-O chain superstructures \cite{Zimmermann2003}. These chains are aligned along the b-axis and exhibit periodic ordering along the a-axis with a periodicity of $ma$. Further investigation using scanning micro-X-ray diffraction has shown that in YBa$_2$Cu$_3$O$_{6.33}$ single crystals, superconductivity emerges within a nanoscale network of oxygen-ordered puddles separated by oxygen-depleted zones \cite{Campi2013}.\\
Regardless of their specific role in superconductivity, the essential heterogeneity plays a critical role in the quantum complexity of the system. Interestingly, it was observed that continuous exposure to X-ray light could induce oxygen ordering and tune the superconducting critical temperature, not by altering the doping content, but by accelerating the mobility and recombination of oxygen interstitials in the chains and striped structures of La$_2$CuO$_{4+y}$ \cite{Poccia2011, Poccia2012, Littlewood2011}. However, X-ray light as an oxygen ordering regulator was effective only when the sample's temperature ranged between 200\,K and 300\,K. Below 200\,K (-73$\,^\circ$C), the Q2 superstructure associated with oxygen ordering could no longer be influenced by X-ray light exposure. At such low temperatures, the oxygen interstitials became immobile, and despite the energy supplied by the continuous X-ray illumination, they could not exhibit any visible oxygen ordering or disordering process.\\
In the case of BSCCO, the mobility of oxygen interstitials is lower because the material is farther from the structural transition compared to in La$_2$CuO$_{4+y}$, making macroscopic phase separation less apparent. However, complex ordering of oxygen interstitials in clusters has been observed in several scanning tunneling microscopy experiments \cite{Ilija2012, Ilija2014, Phillabaum2014, Carlson2015, Song2023}. Subsequent analyses consistently revealed a fractal structure, reflected in the nematic arrangement of the electronic domains, with ultrafast dynamics that are decoupled from lattice directionality.
The analysis shows that nematic clusters emerge from deep within the material and extend to the surface, highlighting significant multiscale order.  Instead, inhomogeneous puddles, a commonly observed experimental feature in cuprates \cite{Ricci2014, Tromp2023, Yu2020, Hayden2024, Osborn2025}, display a slower dynamic, visible especially in systems where the phase separation is larger. In K$_x$Fe$_{2-y}$Se$_2$, time-resolved x-ray photon correlation spectroscopy reveals steady relaxation in paramagnetic domains, while antiferromagnetic regions exhibit intermittent, avalanche-like atomic rearrangements, revealing non-equilibrium dynamics shaped by internal strain \cite{Ricci2020}. In layered nickelates, by contrast, the stability of fluctuating spin stripes is directly tied to the spatial coherence of stripe domains \cite{Ricci2021}. 
The multiscale order and intricate balances between the fractal oxygen ordering distributions and nematic clusters in BSCCO crystal geometry, is also evidenced by bulk supermodulation. Scanning tunneling microscopy reveals that periodic changes in interatomic distances, particularly Cu–O distances, influence the superconducting gap energy \cite{Slezak2008}. These periodic distortions correlate with a supermodulation wavevector, and gap energy variations reflect both the supermodulation and local dopant-induced disorder \cite{Valla2019}. In trilayer BSCCO cuprates, which show even higher superconducting critical temperatures, scanning tunneling microscopy spectra reveal two distinct superconducting gaps, correlating with the inner CuO$_2$ planes and the outer CuO$_2$ planes \cite{Changwei2020}, emphasizing the complex interaction between the supermodulation and intrinsic electronic properties of the BSCCO as well as the importance of the multiscale order \cite{Poccia2011bis}.\\ 
This delicate balance can be observed by exfoliating the crystals at room temperature and encapsulating them to preserve their optimal superconducting properties. Although an optimal $T_c$ of about 90\,K is achieved in 2-unit-cell atomically thin BSCCO crystals, nano X-ray diffraction (nXRD) experiments using a 50\,nm$^2$ beam revealed notable details about the supermodulation wavevector \cite{Poccia2020}. The wavevector, characterized by short- and long-range order superlattice peaks for the atomically thin crystals, remained centered around the bulk average value. However, the shape of its distribution differs from that seen in bulk crystals. This suggests that while the superconducting $T_c$ appears the same and optimal both for the bulk and the atomically thin crystals, the essential heterogeneities in oxygen interstitial ordering at the surface of the BSCCO flakes may be altered by strain or other forms of disorder.\\

\begin{figure*}
  \includegraphics[width=0.9\textwidth]{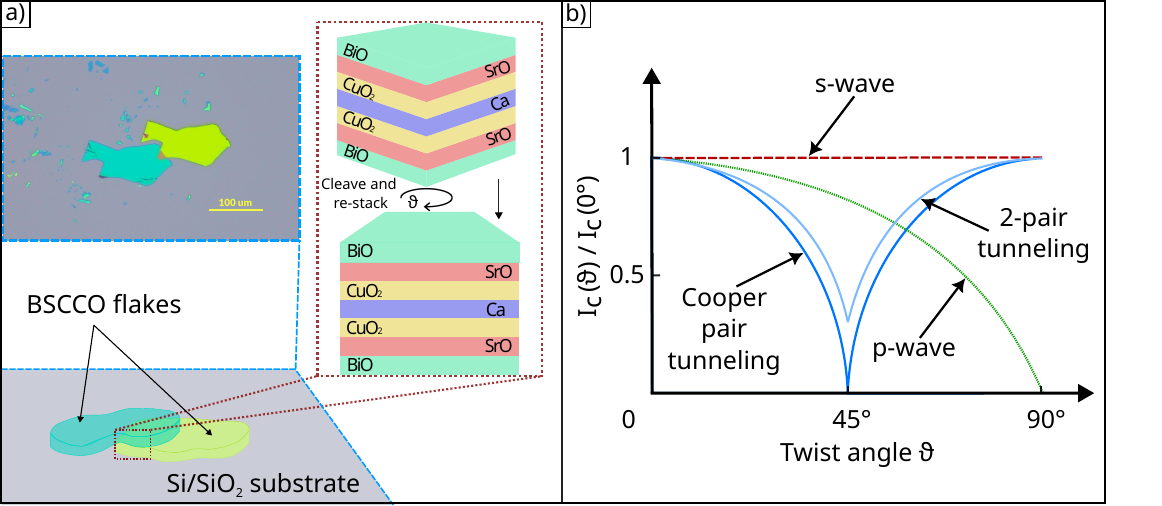}
  \caption{a) Typical optical image of a twisted van der Waals cuprate structure created using the cryogenic stacking technique. The schematic illustrates the junction geometry achieved through the cleaving and restacking of a single flake. b) Expected angular dependence of the critical current, normalized to its value at $\vartheta = 0^\circ$, for s-wave, d-wave 
  (first and second order), and p-wave symmetries.}
  \label{JJstruct}
\end{figure*}

\subsection{Preserving the multiscale order}
The experiments so far suggest that the intricate interplay of spatial heterogeneities, interstitial oxygen ordering, nematic clusters and structural distortion in BSCCO, and cuprates more broadly, is fundamental to their superconducting properties, highlighting the importance of preserving this complexity. Furthermore, as the surface plays a crucial role in forming twisted heterostructures with coherent tunneling characteristics, caution is needed when inferring interface quality solely based on the $T_c$ matching the bulk value. Surface-related heterogeneities, such as oxygen ordering and the coherence of various multiscale orders, may still significantly influence performance. For these reasons, the development of twisted cuprate heterostructures with preserved heterogeneities required innovative techniques to minimize high-temperatures and solvent-induced alterations. Traditional methods for stacking 2D materials, such as twisting graphene and stacking boron nitride, relied on polymers like PPC or PC, requiring high temperatures (higher than 100$\,^\circ$C). On a positive note, Polydimethylsiloxane (PDMS), a silicon-based polymer with a glass transition temperature of -120$\,^\circ$C, emerged as a superior alternative.\\
The unique behavior of PDMS was harnessed to develop a cryogenic stacking technology (CST) for creating twisted cuprate interfaces \cite{Zhao2021}. This approach takes advantage of PDMS’s enhanced adhesion at cryogenic temperatures while simultaneously freezing interstitial oxygen during the fabrication process. The procedure, schematized in Figure (\ref{fabrication}), begins with the mechanical exfoliation of BSCCO crystals onto SiO$_2$/Si substrates in an Ar glovebox. A PDMS stamp mounted on a micromanipulator is then used to cleave the crystal along atomically flat BiO planes by adhering to and rapidly detaching the flakes. Within 90 seconds, the flake on the PDMS stamp is aligned and stacked onto the bottom flake at temperatures between -90$\,^\circ$C and -100$\,^\circ$C. Then the stage is slowly heated to -30$\,^\circ$C, causing the PDMS to lose adhesion and release the top flake onto the bottom one. Electrical contacts are then obtained using a stencil mask technique in an ultra-high or high vacuum environment, eliminating chemical contaminants. The resulting Josephson junctions demonstrated properties consistent with bulk intrinsic junctions without the need of an additional step of annealing, showing a dominant and conventional $d$-wave symmetry dependent on the twist angle \cite{Zhao2023}. Depending on the vacuum level at which the electrical contacts are realized, an encapsulating hexagonal boron nitride (hBN) layer is added immediately after stacking \cite{Lee2023, Martini2023}. Figure (\ref{JJstruct}) shows a typical junction obtained by this method.\\
The studies explore Josephson junctions across various twist angles (0$\,^\circ$ to 180$\,^\circ$), revealing 
physics beyond intrinsic BSCCO junctions. At 0$\,^\circ$, the junctions exhibit electronic quality similar to intrinsic JJs, with hysteretic behavior and dynamic voltage jumps. For intermediate angles (e.g., 29$\,^\circ$ and 39$\,^\circ$), the critical current ($I_c$) and the Josephson coupling ($I_cR_N$) show non-monotonic temperature dependence, linked to BSCCO’s $d$-wave symmetry. At $\theta \approx 45\,^\circ$, where $d$-wave symmetries are maximally mismatched, Josephson coupling weakens but persists, enabling the study of higher-order effects. Signatures of second-harmonic Josephson coupling emerge, evidenced by changes in the current-phase relation, Fraunhofer interference, and fractional Shapiro steps \cite{Zhao2023}.\\
CST was later successfully replicated, allowing the realization of twisted cuprate heterostructures with angular-dependent Josephson coupling while eliminating the need for high-temperature oxygen annealing \cite{Patil2024, Ghosh2024}. In these experiments, the BSCCO flakes were exfoliated and twisted at cryogenic temperatures within an argon-filled glove box, and Au electrodes were realized through a silicon nitride mask using an electron-beam evaporator. Consistent with the first work that utilized the CST, a diode effect was observed at all angles when a perpendicular magnetic field was applied. Moreover, the effect is robust across temperatures, with a maximum asymmetry factor of $60\%$ observed at 20\,K, and the largest diode asymmetry occurred at a 45$\,^\circ$ twist angle, attributed to the interaction between Josephson and Abrikosov vortices.\\
In a separate series of studies, twisted cuprate junctions with different stoichiometries of BSCCO with $T_c$ values of 87\,K \cite{Zhu2023}  and 50–70\,K \cite{Wang2023} were fabricated using the CST. In these cases, after the realization of the junction, it was transferred to a pre-patterned substrate with Ti/Au electrodes for electrical contacts. However, this process may introduce some degree of disorder, as the strain over the interface is challenging to control, which is reflected in the lower Josephson coupling ($I_cR_N$) values measured in the junctions. Moreover, the cleavage and stacking processes occur at temperatures as low as -50$\,^\circ$C, which might be insufficiently low to preserve the essential heterogeneities related to the oxygen ordering. As a result, a significant isotropic $s$-wave pairing component is observed across both optimally and overdoped regimes, with isotropic contributions of $15–20\%$. The studies show an angular dependence of the $I_cR_N$ consistent with mixed $s$- and $d$-wave pairing models.\\
While the microscopic mechanisms in cuprate twisted Josephson junctions remain the subject of ongoing debate, improvements in fabrication techniques, particularly cryogenic stacking, have facilitated consistent measurements of the junction’s angular dependence as a function of the twist angle across multiple studies. This milestone sets the stage for the discussions in the next two sections, which will explore the potential of these junctions in future quantum hardware applications.\\

\section{Theoretical advances in Cuprate twistronics}

\begin{figure*}
  \includegraphics[width=0.9\textwidth]{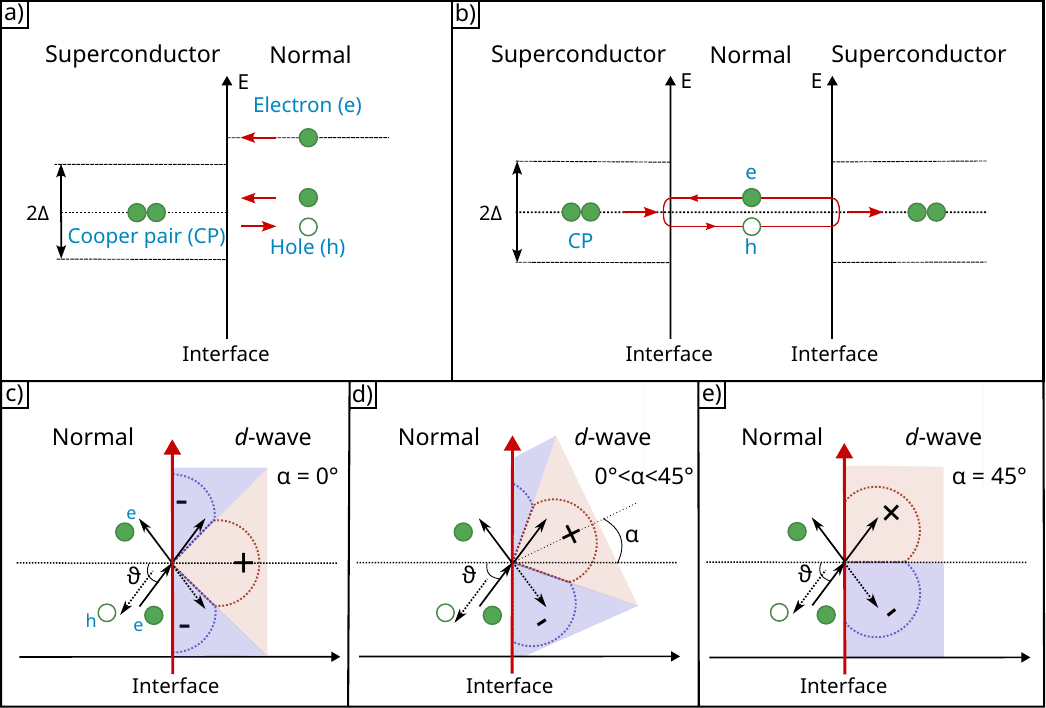}
  \caption{The top panel illustrates the schematic representation of Andreev reflection (a) and the formation of Andreev bound states (b) resulting from multiple Andreev reflections. When an electron from the normal metal with energy below the superconducting gap hits the interface, it is reflected as a hole while a Cooper pair is transmitted into the superconductor. The bottom panel depicts the case of a superconductor with a $d$-wave order parameter (OP). Here, an injected electron from the normal metal undergoes both normal reflection (as an electron) and Andreev reflection (as a hole). The transmitted holelike and electronlike quasiparticles experience different effective pair potentials. Depending on the orientation of OP relative to the interface normal ($\alpha$), the effective pair potential can always have the same sign (c), always opposite signs (e), or a mixed behavior depending on the incident electron’s angle $\vartheta$ (d).}
  \label{Andreev}
\end{figure*}

\subsection{Emerging phenomena at the interface of d-wave superconductors: consequences of order parameter symmetry}
The investigation of tunneling processes in cuprate superconductors began in the 1990s, leading to a wealth of both experimental and theoretical research (see, e.g. Refs. \cite{KirtleyTafuri2007,Kashiwaya1995, Bruder1995, Alff1997, Sigrist1992, Yokoyama2007, Tanaka1995, Tanaka1996, Tanaka1997, Kashiwaya2000}). These studies not only utilized tunneling to confirm the d-wave symmetry of the order parameter but also provided fundamental insights into the physics of high-temperature superconductors. They offered one of the first examples of how bulk superconducting properties can be inferred by examining transport at a material’s surface. In recent years, similar approaches have been applied across a wide range of contexts, from triplet superconductivity to topological systems \cite{Tanaka2021}. Therefore, before discussing the emergence of topological states in twisted van der Waals (vdW) heterostructures, it is pertinent to briefly recall these fundamental studies.\\
Andreev reflection is a fundamental process occurring at normal metal–superconductor (NS) interfaces, where an incident electron from the normal region is retro-reflected as a hole, forming a Cooper pair in the superconductor. This mechanism gives rise to Andreev bound states (ABS), which are localized states at the interface due to multiple Andreev reflections. In unconventional superconductors, particularly those with $d$-wave symmetry, the superconducting order parameter exhibits sign changes depending on the quasiparticle trajectory. This leads to the formation of midgap Andreev states (MAS), a special class of ABS that reside at zero energy due to phase shifts introduced by the sign-changing pairing symmetry. When these states undergo resonant coupling, they form midgap Andreev resonant states (MARS), which strongly influence charge transport, often manifesting as a zero-bias conductance peak (ZBCP) \cite{Kashiwaya1995,Bruder1990, Alff1997}. In certain junction configurations, MAS can form flat-band Andreev states, characterized by a dispersionless energy spectrum, leading to an enhanced density of states at the Fermi level. The general mechanism of Andreev reflection and the reflection and transmission processes at the interface of the N/I/S junction are schematically illustrated in Figure (\ref{Andreev}).\\
The first formula describing the Josephson current in such junctions states that the Josephson current is proportional to $cos(2\alpha)cos(2\beta)$, where $\alpha$ and $\beta$ are the angles between the interface normal and the crystal axes of the left and right d-wave superconductors, respectively \cite{Sigrist1992}. The formula assumes an idealized $d$-wave superconductor/insulator/$d$-wave superconductor configuration and does not consider the influence of MARS \cite{Yokoyama2007}.\\
Theoretical investigations of the role of MAS in tunneling conductance in junctions involving $d$-wave superconductors revealed that they arise from the sign change in the pair potential across the Fermi surface at specific surface orientations \cite{Tanaka1995}. For this reason, ZBCPs are typically observed in $ab$-plane tunneling configurations when the tunneling direction is misaligned with the crystalline axes of the superconductor and are considered strong evidence of $d$-wave pairing symmetry. Additionally, the Bogoliubov-de Gennes equation for the $d$-wave superconductor was extended, showing that the midgap states are further influenced by factors such as the barrier height, interface orientation, and tunneling direction \cite{Tanaka1997}. In a twin work, the local density of states of quasiparticles near the interface of $d$-wave superconductor was studied, demonstrating the existence of zero-energy states (ZES) that contribute to zero-energy tunneling conductance peaks \cite{Tanaka1996}. Due to these ZES, for low-conductance $d$-wave junctions, the maximum Josephson current $I_{c}$ is enhanced at low temperatures \cite{Kashiwaya2000}.\\
The anomalous temperature behavior, absent in $s$-wave junctions, was also discussed in twisted cuprate vertical heterostructures, where the critical current increases with temperature up to a point before decreasing, reflecting the interplay between nodal and antinodal contributions. The longstanding assumption of $s$-wave symmetry for the c-axis gap was revised, introducing a more accurate equation that suggests strong evidence for $d$-wave gap symmetry \cite{Talantsev2024}. Experimental data on twisted cuprate flakes with non-monotonic temperature dependence of critical current $I_c$ as a function of twist angle show behaviors consistent with $d$-wave superconductivity, including angular dependencies and gap suppression at specific angles due to symmetry \cite{Tummuru2022}, further confirming the nodal properties of the gap. \\
A first comprehensive discussion about $d$-wave Josephson junctions highlighting the possible completely new phenomena emerging from the anisotropy of the SOP can be found in Ref. \cite{Sigrist1998}. Near surfaces and interfaces, such as Josephson junctions, localized states can form where time-reversal symmetry (TRS) is broken. When TRS is broken, spontaneous supercurrents appear, leading to unusual quantum effects such as fractional flux vortices and phase-slip phenomena in Josephson junctions. In these systems, ABS that emerge at the surface can exhibit flat or linear dispersion, stabilized by topological invariants \cite{Sato2011}. The existence of these zero-energy surface states results from topologically protected nodal structures, described by Chern numbers in TRS-breaking systems and $\mathbb{Z}_2$ invariants in TRS-preserving systems \cite{Kobayashi2015}. These topological properties predict the presence of ABS and their potential to host Majorana fermions, particles of interest for quantum computing. Moreover, in TRS-breaking superconductors, ABS behave similarly to quantum Hall edge states, leading to spontaneous edge currents. These findings help classify cuprates as topological superconductors, highlighting their connection to modern topological materials.\\
In 2021, it was theoretically predicted that topological superconductivity in cuprates could be precisely tuned by twisting $d$-wave superconductors, such as BSCCO, at angles near 45$\,^\circ$ \cite{Can2021}. The interplay between the twisted layers induces a secondary $d_{xy}$ like superconducting order in each layer, creating a combined $d+id'$ state. This state is fully gapped and topological, appearing very close to the bulk critical superconducting transition temperature and with a Chern number $C=\pm2$. An independent group validates these theoretical results, reporting that above the critical temperature of the chiral topological superconductors, formed by twisting cuprate layers, two vestigial phases arise: a charge-4$e$ superconducting phase, involving the pairing of Cooper pairs into quartets, and a chiral metal phase, characterized by restored time-reversal symmetry \cite{Liu2023, Liu2023bis}. Using symmetry-based theoretical modeling, renormalization group analysis, and Monte Carlo simulations, the study provides a detailed phase diagram and practical framework for realizing charge-4$e$ superconductivity.\\
Low twist angles in twisted cuprate superconductors, modeled as square lattices, have been studied in scenarios where symmetry prevents direct hopping between adjacent sites, leading to the formation of decoupled sublattices \cite{Volkov2023}. Breaking this symmetry, for instance, by applying an electric or magnetic field, introduces tunable hopping terms, allowing control over hopping ratios crucial for investigating correlated electron behavior \cite{Eugenio2024}. An in-plane magnetic field can generate a periodic lattice of topological domains with alternating Chern numbers. The chiral edge modes of these domains form low-energy bands, detectable through scanning tunneling microscopy, and can lead to a quantized thermal Hall effect at low temperatures.\\
The transition near a 45$\,^\circ$ twist angle where the Josephson current-phase relationship shifts from $2\pi$ to $\pi$ periodic is caused by second-order Cooper pair cotunneling processes \cite{Volkov2025}. At 45$\,^\circ$, first-order Josephson tunneling vanishes by symmetry, but higher-order cotunneling processes dominate, giving rise to a second-harmonic current-phase relationship. Tunneling form factors reveal unique pairing and energy-phase relations, with notable double-Cooper-pair tunneling dominating \cite{Song2022}. Experimental signs of this transition are expected to include magnetic fields causing altered Fraunhofer oscillations with extra nodes, as well as radiofrequency drives resulting in fractional Shapiro steps. Theory also suggests that time reversal symmetry broken states can persist in structures with multiple monolayers, although the topological gap decreases exponentially with thickness \cite{Song2022}. Interestingly, although the dominant SOP is $d$-wave, as confirmed by experiments, secondary order parameters can appear, with secondary instabilities like $s$-wave order emerging and adding complexity \cite{Volkov2023bis}.\\
Calculations on the Hubbard model for twisted cuprates explore the impact of interlayer doping asymmetry on time-reversal symmetry-breaking states in bilayer systems \cite{Belanger2024, Belanger2024bis}. Using the variational cluster approximation, twisted bilayers at $\theta = 53\,^\circ$ were studied with an interlayer bias simulating doping differences. The results show that chiral superconductivity is highly sensitive to this bias, while moving closer to a 45$\,^\circ$ twist does not significantly enhance symmetry breaking due to the similar nodal structures of competing states. Strong interlayer tunneling is essential for stabilizing the symmetry-breaking phase, which remains robust against small variations in model parameters but is weakened by reduced hopping amplitudes, affecting the reversal symmetry breaking phase.\\

\subsection{Josephson diode effects in twisted cuprates}
Such systems have been proposed to be particularly favorable for the observation of the so-called Josephson diode effect. It appears in superconducting systems where electrical current flows more easily in one direction than the other, similar to a semiconductor diode but without resistance. Differently from standard cases, where current can move symmetrically in both directions, a Josephson diode exhibits nonreciprocal superconducting transport, meaning it allows supercurrent to flow preferentially in one direction while suppressing or even blocking it in the opposite direction. Although the superconducting diode effect is common in thin film configurations (see e.g. Ref \cite{Hou2023}) and fluxons in Josephson junctions with inhomogeneities cause asymmetries in I-V curves, a combination of time-reversal symmetry (TRS) breaking and inversion symmetry breaking may create more favorable conditions for the diode effect to arise in such systems. Josephson diode effects are considered in proposals for superconducting electronics, enabling high-frequency rectification, filters, and energy-efficient devices, all operating without external magnetic fields \cite{Guarcello2024, Yerin2024}. It was demonstrated that under slow driving, a junction exhibiting time-reversal symmetry-breaking effects can enter a regime where it behaves as an ideal superconducting diode, carrying zero-voltage direct current only in one direction and achieving an ideal superconducting diode efficiency of $\eta=\pm1$ \cite{Seoane2024}. This is characterized by a regime where the critical current asymmetry becomes maximized. With fast driving instead, the diode efficiency can still be modulated but does not reach the ideal regime. \\
Inspired by experiments on cuprate twistronics, the focus is placed on the Josephson diode effect and its behavior, categorized into two types \cite{Volkov2024}. The dynamical Josephson diode effect arises from spontaneous time-reversal symmetry breaking near a 45$\,^\circ$ twist angle, while the thermodynamic Josephson diode effect results from explicit time-reversal symmetry breaking, such as the influence of magnetic fields. The dynamical effect relies on the bistability of the Josephson free energy, allowing for trainable diode polarity through current sweeping. In contrast, the thermodynamic effect has a fixed polarity and remains stable even in strongly damped systems. Josephson diode effects in $d$-wave pairings typically require spin-orbit coupling and an external magnetic field. However, if $d+id$ or $d+is$ pairings form, they inherently break time-reversal symmetry, making Josephson diode behavior possible even without a magnetic field \cite{Vakili2024}. Interestingly, in $d+id$ pairings, chiral edge states play a crucial role in the Josephson diode effect, producing distinctive signatures, whereas $d+is$ pairings exhibit no such edge-state contributions.\\
In chiral $d$-wave superconductors, the magnitude and distribution of the edge currents depend on the type of edge or domain wall in the system. These edge currents, although smaller than those expected in chiral $p$-wave superconductors, are nonzero and could be detected experimentally \cite{Holmvall2025}.  Very recently, it was shown that edge currents are present in the chiral $d+id$ phase, emerging from the spontaneous time-reversal symmetry breaking at the interface of twisted BSCCO bilayers \cite{Pathak2024}. The edge currents should be strong enough to be observed experimentally \cite{Sigrist1995, Bailey1997}, furthering the understanding of chiral $d$-wave superconductivity and its potential applications.\\
An intriguing proposal explores the interplay between the intrinsic pairing properties of cuprate bilayers and the Rashba effects induced by a specially designed substrate, a synergy confirmed through density functional theory simulations. \cite{Margalit2022}. The induced superconducting gap in the substrate demonstrates that the heterostructure can support chiral Majorana modes within an experimentally observable temperature range. The findings suggest that this heterostructure, leveraging the high pairing energy of cuprates and the distinct $(d+id)$-wave symmetry, provides a viable path toward realizing chiral topological superconductors.  The proposed system combines twisted BSCCO cuprates with the topological insulator Bi$_2$Se$_3$, utilizing spin-orbit coupling to host Majorana zero modes. Analytical Bogoliubov–de Gennes (BdG) formalism and numerical lattice simulations confirm their presence in vortex cores. Key advantages include high-temperature operation up to the cuprate’s native $T_c$, strong Majorana protection due to the short coherence length and limited quasiparticle states, and a larger induced superconducting gap compared to conventional $s$-wave systems. The combination of a twisted cuprate bilayer with a two-dimensional topological insulator enables the $d+id$ phase to generate a uniform Dirac mass in the helical edge states, while a Zeeman field induces contrasting mass terms along different edges \cite{Mercado2022}. At critical Zeeman field values, mass domain walls form at the corners, giving rise to Majorana zero modes as mass-kink excitations. The interplay between $d+id$ pairing and the Zeeman field gaps out edge states, with alternating Dirac mass signs on adjacent edges, resulting in a single Majorana corner mode at each corner \cite{Li2023}.\\

\subsection{Influence of disorder and dissipative effects}
Early studies on the role of disorder in junctions between $d$-wave superconductors examined the proximity effect, revealing that the superconducting order parameter can be suppressed near a boundary depending on the crystal orientation, significantly affecting the dc Josephson current \cite{Barash1995}. Additionally, quasiparticle tunneling was analyzed, highlighting unique I-V characteristics such as zero-bias anomalies and conductance jumps linked to the anisotropy of the $d$-wave gap. However, since most configurations were based on the epitaxial growth of in-plane junctions, advancements in current technologies now allow for the mitigation of certain disorder effects. Nevertheless, disorder remains a crucial factor, necessitating precise techniques for fabricating these interfaces. Indeed, interface roughness affects momentum conservation, reducing the critical current and influencing its angular dependence. This could explain why only very recently a clean $\cos(2\theta)$ angular dependence of the critical current was observed in vertically twisted cuprate junctions. Furthermore, strong disorder suppresses the $d+id$ phase, thereby restoring time-reversal symmetry.\\
The role of disorder appears to be more subtle. The second order coupling is the dominant one near $\theta = 45\,^\circ$, however, the overall critical current remains unexpectedly large, surpassing the much smaller values predicted by intrinsic mechanisms. This anomalously large second-order coupling observed in experiments is suggested to arise from extrinsic mechanisms. Specifically, intrinsic inhomogeneities induce a random local first-order coupling, which in turn generates a global second-order effect favoring time-reversal symmetry breaking \cite{Yuan2023}. Potential sources of such intrinsic disorder include the presence of electron nematic domains, as supported by scanning tunneling microscopy studies, and structural heterogeneities, described by puddles models \cite{Alvarez2005, Velasco2023, Caprara2013}. Ultimately, cryogenic technology has been developed to preserve the configuration of these intrinsic domains in BSCCO, with such inhomogeneities locally mixing superconducting order parameters. On the other hand disorder may, to some extent, play a beneficial role in the topological superconductivity of twisted cuprate heterostructures. Incoherent tunneling mediated by impurities in spacer layers is investigated using a continuum model and numerical lattice simulations \cite{Haenel2022}. The findings demonstrate that while disorder preserves crystal symmetry on average, it can induce a gapped, time-reversal symmetry-breaking, and topologically nontrivial phase near $\theta = 45\,^\circ$, highlighting the critical role of disorder in stabilizing the gapped topological state. \\
The robustness of the electronic properties of the twisted cuprates are discussed using a moiré cell-based bilayer model, where the impact of single-site and dimer impurities on both topologically trivial and nontrivial superconducting states is explored \cite{Xie2024}. The gapped, time reversal symmetry broken and topologically nontrivial phase near $\theta = 45\,^\circ$ is highly resilient to single scalar-potential impurities but sensitive to interlayer impurity dimers, which can enforce or prevent subgap quasiparticle state formation. Additionally, single impurities of different types (e.g., positively charged or magnetic) exhibit distinct signatures on the local density of states, as revealed by BdG theory.\\
In the view of possible application of cuprate twistronics, dissipative effects and vortex dynamics play a crucial role. In general, the physics of the vortex states in high-temperature superconductors is complex and might be difficult to control unless specific experimental protocols are realized \cite{Blatter1994}. Vortices behave like elastic lines and exhibit dynamic behaviors influenced by disorder and the periodic potential of the superconducting lattice \cite{Ioffe1987}. This leads to thermally activated vortex motion or vortex creep, where vortex movement under applied currents is highly nonlinear. Unlike typical particles, whose velocity is linearly proportional to force, vortex motion involves a friction coefficient that depends exponentially on temperature and applied force. Below the depinning temperature, this nonlinearity lead to slow and irregular motion at low currents. Additionally, the time it takes for vortices to escape their initial positions follows a statistical distribution influenced by the pinning barrier \cite{Feigelman1989,Marchetti1994,Vinokur1996}.\\
For twisted bilayer superconductors, a recent theoretical work predicts skyrmionic vortex states using Ginzburg-Landau theory \cite{Cadorim2024}. These states evolve from composite vortex lattices to extended skyrmionic chains as twist angles approach $\theta = 45\,^\circ$,  demonstrating how specific vortex configurations can directly reveal the underlying topological phase. Indeed, an additional study has predicted the formation of quadruply quantized coreless vortices stabilized by fractional vortices and domain walls separating chiral components \cite{Holmvall2023}. These coreless vortices show unique behaviors under opposite magnetic fields, breaking rotational and axial symmetries and producing measurable differences in the local density of states and magnetic moments. These features, distinct from conventional Abrikosov vortices, provide experimental evidence for time-reversal symmetry breaking, chiral superconductivity, and the Chern number, accessible through techniques like scanning tunneling spectroscopy and magnetometry.\\

\subsection{Multilayer structures}
Building on the insights gained from bilayer systems, focus has also been extended to multilayer structures, which introduce additional degrees of freedom and richer physical phenomena. By stacking multiple cuprate layers with controlled twist angles, novel interference effects emerge, deepening the understanding of how twisting influences nodal superconductors. Twisted nodal superconductors in multilayer systems were investigated, focusing on two distinct high-symmetry stacking configurations: alternating twists and chiral twists \cite{Tummuru2022bis}. In the alternating twist configuration, adjacent layers alternate between positive and negative twist angles, producing features similar to twisted bilayers, such as time-reversal symmetry-breaking superconductivity near 45$\,^\circ$ twist angles. However, in odd-layer systems, this configuration results in a gapless spectrum due to the presence of a decoupled monolayer sector. In contrast, chiral twist stacking, where all layers are twisted in the same direction, leads to fully gapped chiral topological superconductivity with chiral Majorana edge modes. Through a BCS mean-field approach, the study finds that alternating twist stacking enhances tunneling and leads to larger magic angles, where Dirac nodes merge into quadratic band touchings, promoting $d+id$ superconductivity. Meanwhile, in chiral twist stacking, cubic band touchings with a divergent density of states trigger gap-opening instabilities, stabilizing $d+id$ phases over a range of twist angles.\\
A complementary analysis, utilizing Cartan symmetry classifications and focusing on the superconducting Altland-Zirnbauer (AZ) symmetry classes, investigates how perturbations in multilayers can lead to the emergence of topologically nontrivial superconductors characterized by a $\mathbb{Z}_2$ invariant \cite{Lucht2024}. When a Josephson spin current is applied between layers, it generates a spin-phase difference, inducing a perturbation that opens a gap in the spectrum and leads to the formation of a helical topological phase, particularly in triplet superconductors. This perturbation breaks time-reversal symmetry and triggers a $\mathbb{Z}_2$ gap, essential for the formation of topological superconductivity. However, for singlet superconductors, such a mechanism fails, requiring the introduction of a triplet order parameter to achieve a nontrivial $\mathbb{Z}_2$ invariant. The analysis further shows that the topological properties of these systems are affected by the number of layers and the arrangement of twists. For even-layered systems, the $\mathbb{Z}_2$ invariant is trivial, while odd-layer systems can exhibit nontrivial topological indices, depending on the specific twist configurations. \\

\begin{figure*}
  \includegraphics[width=0.8\textwidth]{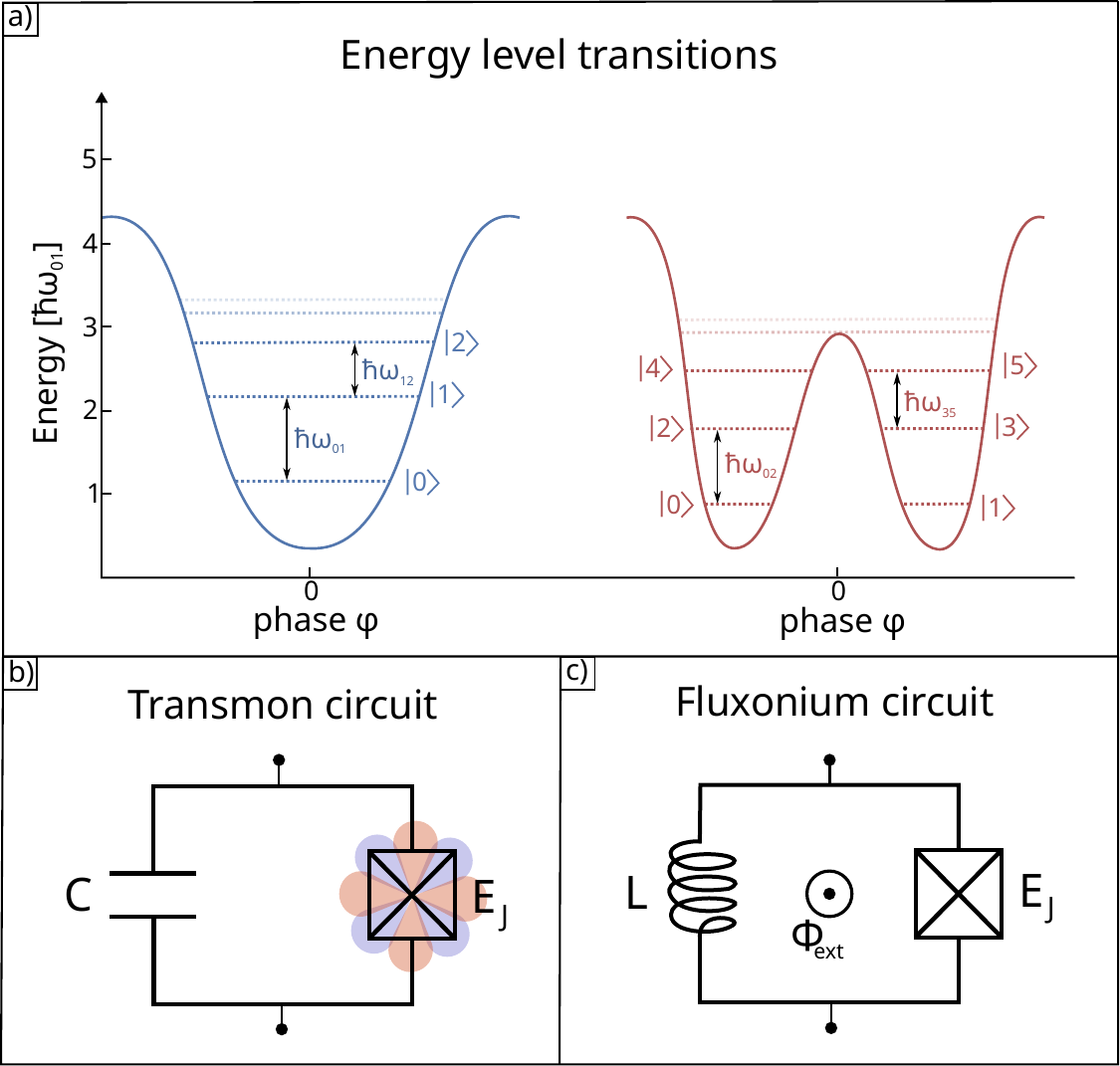}
  \caption{a) Schematic representation of the qubit concept. On the left a cos($\varphi$) potential, typical of a transmon system, and on the right a double-well potential, peculiar to systems such as the fluxonium. The two lowest energy states $\lvert0\rangle$ and $\lvert1\rangle$ form the computational basis. The energy level spacing is anharmonic, enabling precise quantum control. On the bottom, equivalent circuits of the two main qubit types: transmon-qubit (b) and flux-based qubit (c). Flux-based qubits store quantum information in magnetic flux through a Josephson junction loop, making them less sensitive to electric fields and reducing crosstalk. They allow precise control via magnetic flux tuning but are susceptible to magnetic noise. A transmon qubit uses a Josephson junction with a large shunt capacitor to reduce charge noise sensitivity but becomes more susceptible to microwave electric noise from surface dielectrics. If the Josephson junction in a transmon configuration is realized with a twisted cuprate junction, such as the one schematically depicted in b), the design is referred to as a \textit{flowermon}. The symbol $C$, $L$, $E_J$, $\phi_{ext}$ represent, respectively, the capacitor, the inductance, the Josepshson junction and the external magnetic field of the circuit.}
  \label{qubits}
\end{figure*}

\section{Cuprate twistronics for quantum hardware}

\subsection{Qubit classification}
Superconducting qubits are the fundamental building blocks of many quantum computers. They are electronic circuits that mimic artifical atoms, storing quantum information as quantized energy levels and allowing precise manipulation through microwave signals \cite{Devoret2013}. At the core of a superconducting qubit lies a high-coherence microwave LC circuit with naturally quantized energy levels, featuring a Josephson junction. The Josephson junction introduces essential nonlinearity, creating unevenly spaced energy level transitions. This enables the isolation and control of specific transitions, which are crucial for quantum operations. The physical components of superconducting qubits are constructed using superconducting materials such as aluminum (Al), niobium (Nb), tantalum (Ta), niobium nitride (NbN), and niobium titanium nitride (NbTiN). The Josephson junction itself is typically fabricated from an Al/AlO$_x$/Al structure, as the native aluminum oxide layer offers exceptional quality and stability for quantum operations. \\
Superconducting qubit designs can very roughly be categorized into two types: (i) \textit{charge-based qubits}, predominantly represented by the transmon qubit design \cite{Koch2007}; and (ii) \textit{flux-based qubits}, such as the flux qubit \cite{Mooij1999,Yan2016} and fluxonium \cite{Manucharyan2009}. In a transmon qubit, a Josephson junction is shunted by a large capacitor, which reduces sensitivity to charge fluctuations. However, the large capacitor size (approximately 100$\,\mu$m) increases susceptibility to microwave electric noise, primarily originating from imperfect surface dielectrics near the capacitor. This issue has driven extensive efforts to eliminate oxide layers and improve substrate quality. Additionally, transmon qubits can unintentionally interact through shared electric fields, further complicating multi-qubit systems.\\
Flux-based qubits, in contrast, encode quantum information in the magnetic flux passing through a loop of Josephson junctions. These designs are less sensitive to electric fields and exhibit reduced crosstalk in multi-qubit architectures. By tuning the magnetic flux through the loop, flux qubits offer greater control but also introduce vulnerability to magnetic noise. To mitigate this, flux qubits are often operated at specific bias points, such as at full or half-integer magnetic flux quanta through the loop, where they are less sensitive to magnetic field fluctuations. However, this strategy limits tunability and increases the complexity of the system, requiring continuous adjustment and monitoring of the external magnetic flux. \\
Figure (\ref{qubits}) provides an overview of the working principles of a superconducting qubit, along with the equivalent circuits for its two main categories. As a leading platform for quantum computing, superconducting qubits are constantly evolving to enhance coherence, noise resilience, and scalability. However, there is a growing consensus that emerging 
types of JJs can pave the way to innovative architectures and alternative hardware solutions. \\

\begin{figure*}
  \includegraphics[width= 0.8\textwidth]{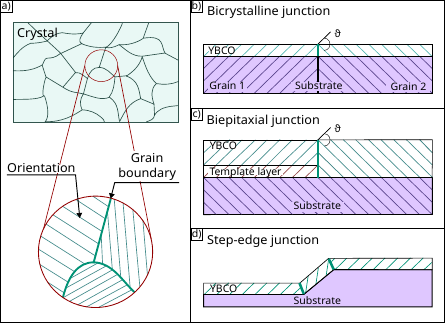}
  \caption{a) Schematics of grain boundaries in a generic crystal; on the right, types of grain boundaries Josephson junctions: b) in bicrystalline junctions a film is grown on a bicrystalline substrate with a grain boundary of a certain configuration. In this way, the grain boundary in the film is the same as the substrate; c) in biepitaxial junctions, the process used relies on changes in the orientation of high-$T_c$ films caused by epitaxial growth on patterned template layers. The in-plane rotation of the high-$T_c$ films depends on the structure of the underlying layers, leading to the formation of rotational grain boundaries at the template edges; d) step-edge junctions show two grain boundaries which are grown at the bottom and the top of a substrate step.} 
  \label{Grain boundaries}
\end{figure*}

\subsection{Grain boundaries Josephson junctions as a platform for quantum devices}
\label{sec:GBJJs}
A comprehensive review has gathered most experiments supporting a predominantly $d$-wave order parameter \cite{Tsuei2000}. Penetration depth, specific heat, and angle-resolved photoemission spectroscopy all reveal gap anisotropy with nodes in the superconducting state. Experiments like superconducting quantum interferometry and Josephson junctions demonstrate phase shifts consistent with the hypothesis. The nature of this pairing symmetry is proposed to be important for potential applications in quantum technologies \cite{Tsuei2002, Tsuei2004}.\\
The first high-temperature superconductor (HTS) qubit was designed to be quiet, leveraging the unconventional $d$-wave symmetry of high-$T_c$ superconductors to construct bistable qubits that operate without external magnetic fields or current biases, reducing environmental interference \cite{Ioffe1999}. The design utilizes double-periodic energy-phase characteristics of the junctions, with minima at phase shifts of 0 and $\pi$. To enhance qubit stability and control, it was also proposed to integrate these junctions into five-junction loops, which create effective $2\pi$-periodic junctions with tunable phase shifts. However, experimentally, HTS junctions with cuprates immediately showed enormous challenges.\\
A grain boundary junction is the interface where two crystalline regions in a polycrystalline material meet, disrupting the crystalline order (Figure \ref{Grain boundaries}) \cite{Hilgenkamp2002}. In cuprates, these boundaries are particularly significant due to their highly anisotropic crystal structures and short coherence lengths, which make them highly sensitive to atomic misalignments. As a result, grain boundaries can act as weak links, impacting the material's electrical and superconducting properties. Grain boundaries are categorized into bicrystalline \cite{Mannhart2001}, biepitaxial \cite{Tafuri1999, Char1991, DiChiara1997}, and step-edge junctions \cite{Edwards1992}, enabling precise study and engineering of these interfaces (Figure \ref{Grain boundaries}). Advanced techniques, such as irradiation and template-layer processes, are used to tailor grain boundaries for specific applications \cite{Tafuri2005}. At low misorientation angles, grain boundaries feature dislocation arrays, while high-angle boundaries form continuous disordered layers \cite{Hilgenkamp2002}. Phenomena such as Josephson behavior or flux flow dominate depending on the misorientation angle and temperature, with factors like structural distortions, stoichiometric deviations, order parameter symmetry, and electronic effects (e.g., band bending) further shaping boundary behavior \cite{Tafuri2005, Hilgenkamp2002}.\\
In a set of experiments based on the grain boundary junctions made with yttrium based high temperature cuprate superconductors, the critical current is observed to vary with the angle of the grain boundary, showing minima at specific orientations (0$\,^\circ$, 34$\,^\circ$, and 90$\,^\circ$), consistent with the Sigrist-Rice phenomenological model, and confirming the angular dependence  expected for a $d$-wave pairing state \cite{Lombardi2002}. The coherence of the $d$-wave order parameter extends beyond the characteristic length scales of interface imperfections, attributed to low-barrier transmission probabilities in the junctions. Driving this research on the in-plane cuprate grain boundaries junctions, besides the demonstration of dominant pairing symmetry, it was the prospect of the application of cuprates Josephson junctions for quantum hardware.\\
Further investigation in this direction examined the quantum dynamics in high-temperature cuprate Josephson junctions, specifically through symmetric 45$\,^\circ$ YBa$_{2}$Cu$_{3}$O$_{7-x}$ (YBCO) grain boundary junctions, revealing several temperature-dependent anomalies in the current-phase relationship \cite{Il'Ichev2001}. The critical current exhibits a local minimum at a specific temperature of $T^\ast$, around which the first harmonic of the Josephson current (I1) undergoes a sign change, indicating a phase inversion in the current phase relationship. Simultaneously, the second harmonic (I2) becomes comparable to or even surpasses I1, leading to a non-standard current phase relationship. At low temperatures, the junction develops a doubly degenerate ground state, with two distinct energy minima corresponding to different phase differences. At the time, the challenge remains the junction transparency and reduction of detrimental disorder effects.\\
Using a bi-epitaxial technique, YBCO grain boundary junctions that formed bistable systems were fabricated. The junctions were modeled using an RCSJ framework that incorporates the effects of stray capacitance and the kinetic inductance of the SrTiO$_{3}$ (STO) substrate. Experiments showed clear quantum state transitions under microwave irradiation, with a quality factor (Q) of $\sim$40, comparable to low-temperature superconductors despite the presence of nodal quasiparticles. The junction design was further improved by reducing stray capacitance and controlling dissipation, demonstrating a transition from macroscopic quantum tunneling below 135\,mK to diffusive Brownian motion above this temperature \cite{Longobardi2012}. In the following years, YBCO grain boundary junctions with widths down to 600\,nm were fabricated using electron-beam lithography and ion-beam etching. (La$_{0.3}$Sr$_{0.7}$)(Al$_{0.65}$Ta$_{0.35}$)O$_{3}$ (LSAT) substrates reduced parasitic capacitance compared to standard STO substrates, enhancing performance \cite{Stornaiuolo2013}. The devices exhibited low critical current densities due to oxygen depletion in the grain boundary regions, which still constitute a source of detrimental disorder in these junctions. However, the Josephson energy was calculated to be significantly lower than typical high temperature superconductor devices, allowing the exploration of poorly studied dynamical regimes.\\
In BSCCO, the macroscopic quantum tunneling was observed at relatively high crossover temperatures ($\sim$1\,K), significantly higher than in conventional $s$-wave superconductors ($\sim$300\,mK) \cite{Inomata2005}. This behavior was attributed to the high plasma frequency of intrinsic Josephson junctions, underscoring the potential of BSCCO Josephson junctions for ultrafast quantum electronics \cite{Jin2006}.\\

\begin{figure*}
  \includegraphics[width=0.6\textwidth]{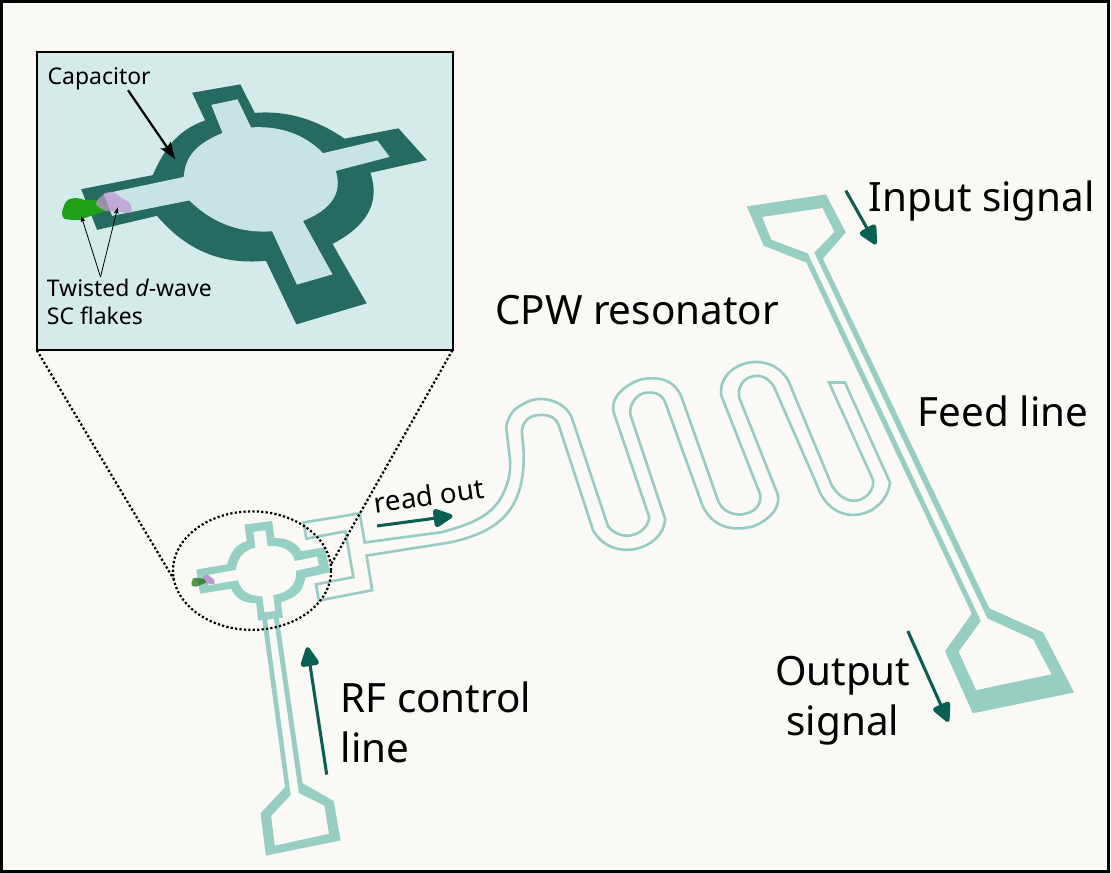}
  \caption{Schematic of a realistic setup for controlling and computing with a qubit based on the \textit{flowermon} design. The inset highlights the near 45$^\circ$ twisted cuprate junction, connected in parallel with a capacitor to form the \textit{flowermon}.}
  \label{res}
\end{figure*}

\subsection{New qubits architectures based on anisotropic order parameter}
Twisted cuprate junctions can be seen both as artificial materials hosting unconventional topological superconducting states and as promising structures for the development of quantum processing devices. The early experiments on grain boundary junctions described in the previous section demonstrated that nodal quasiparticle tunneling may significantly hamper the performance of superconducting quantum devices based on these types of junctions. On the other hand, theoretical works indicate that Josephson junctions twisted along the c-axis could offer enhanced protection against quasi-particle tunneling. These early works focused mostly on macroscopic quantum tunneling (MQT) utilizing functional integral methods to derive the MQT rate for $d$-wave symmetry Josephson junctions twisted along the c-axis, rather than the ab-plane used in grain boundary junctions \cite{Kawabata2004}.\\
It was demonstrated that untwisted junctions suffer from super-Ohmic dissipation caused by nodal quasiparticle tunneling. In this configuration, the alignment of the superconducting nodes facilitates quasiparticle tunneling, even at low temperatures, hindering the observation of MQT even in clean samples. By contrast, twisting the junctions inhibits node-to-node tunneling, reducing dissipation \cite{Yokoyama2007bis}. Additionally, the standard deviation of the switching current distribution and the crossover temperature from thermal activation to quantum tunneling vary with the cosine of twice the twist angle.\\
Recently, van der Waals (vdW) assembly and a deeper understanding of cuprate materials and superconducting quantum circuits led to 
superconducting qubit designs based on twisted vdW cuprate junctions \cite{Brosco2024, Patel2024}. The \textit{flowermon} design, introduced in Ref.\cite{Brosco2024}, integrates a vdW cuprate junction with a twist angle close to  45$^\circ$, combined with a shunting capacitor in a circuit QED setup \cite{Blais2021, Devoret2013}. The twisted $d$-wave nature of the order parameter leads to a strong suppression of single Cooper pair tunneling and makes the second-harmonic Josephson tunneling, $\propto \cos(2 \varphi)$, the dominant contribution to the Josephson energy.  This enables the creation of an intrinsically protected qubit.\\
The idea of using two minima of a $\cos(2\varphi)$  potential to define two inequivalent quantum states for realizing a protected qubit has been explored in various works \cite{Ioffe1999,Kitaev2006, Gladchenko2009, Douçot2012, Brooks2013, Bell2014, Smith2020, Larsen2020, Gyenis2021, Dodge2023, Ciaccia2024,Blatter2001}. The $\cos(2\varphi)$ junction allows only tunneling of pairs of Cooper pairs, thus separating the computational space into states with an even or odd number of Cooper pairs and yielding topological protection against dielectric and charge fluctuations. Several platforms have been considered to use the two-Cooper-pair tunneling property for protected qubit encoding, such as the rhombus \cite{Blatter2001,Douçot2012,Larsen2020,Ciaccia2024}, the KITE \cite{Smith2020}, and the semi-superconducting qubits \cite{Larsen2020, Ciaccia2024}.
The \textit{flowermon} design offers the advantage that the protection is built into the many-body wave-function and does not require external fields or circuit engineering, which introduce disorder and fluctuations. The idea of \textit{flowermon} is therefore related to Refs.\cite{Blatter2001, Ioffe1999, Blais2000}, which first explored the suppression of tunneling in $d$-wave based Josephson junctions to realize superconducting qubits. Interestingly, employing modern approaches to quasi-particle-induced decoherence, it was possible to demonstrate that under appropriate conditions, large twisting angles suppress the effects of these kinds of processes on relaxation and dephasing in the \textit{flowermon} qubit, reminiscent of what happens in MQT. In Figure (\ref{res}), a possible scheme of integration of a \textit{flowermon} qubit in a circuit QED architecture with a readout and control resonator is presented.\\
Developing from it, a theoretical proposal suggests that by integrating two twisted junctions into a superconducting quantum interference loop threaded with magnetic flux, the system can be tuned to various regimes by adjusting the external flux and twist angle \cite{Coppo2024}. These regimes include a symmetric \textit{twist-based} double-well potential, a \textit{plasmonic} potential, and a \textit{flux-biased} double-well potential. Each regime affects the system’s sensitivity to external noise, with the twist-based regime offering significant protection against both charge and flux noise. The study also introduces a supersymmetric quantum circuit where the spectrum has one non-degenerate ground state and pairs of degenerate excited states.\\
An alternative design based on a Josephson junction between a $d$-wave superconductor (e.g. such as the BSCCO) and an $s$-wave superconductor, called \textit{d-mon}, is robust against offset charge fluctuations, like conventional transmons \cite{Patel2024}. Operating in a regime with fully gapped quasiparticles minimizes the decoherence risks typically associated with nodal quasiparticles in $d$-wave systems. The qubit's $\pi$-periodic Josephson free energy yields a parity-protected qubit supporting strong anharmonicity. Additionally, the design can be expanded in a split \textit{d-mon} variant with a three-junction design that allows fine-tuning of parameters via magnetic flux, enhancing its versatility \cite{Patel2024}.\\

\section{Outlook}
The field of cuprate twistronics, manipulating twisted van der Waals layers of cuprate superconductors, is experiencing a revival driven by the development of the cryogenic stacking technique in an inert atmosphere that preserve the pristine structure of these fragile materials. In parallel, a surge in theoretical predictions is uncovering new opportunities in both applied and fundamental physics. This progress has reignited interest in van der Waals cuprates for superconducting circuits, opening 
unexplored avenues for quantum hardware. A notable example is the successful integration of BSCCO flakes into niobium superconducting resonators, demonstrating strong coupling, low dissipation, and notable modifications in resonator behavior \cite{Jin2025}.\\
However, current nanofabrication techniques face significant challenges in producing stable, ultra-thin, and clean cuprate heterostructures or nanowires, as these materials tend to degrade during complex circuit processing. To address the delicate issue of making reliable electrical contacts on ultra-thin 2D materials, 
innovative experimental approaches have been proposed \cite{Liu2018, Yang2023, Saggau2023, Martini2023bis, Shokri2024}. One of them suggests creating SiN$_x$ membranes patterned with circuits and stacking them onto cuprate van der Waals layers at cryogenic temperatures, preserving their properties \cite{Saggau2023,Martini2023bis}. This approach separates circuit fabrication from the manipulation of the delicate structures, avoiding the use of polymers, solvents, or high temperatures. The method has demonstrated both the reproduction of Hall-effect data and the creation of thermoelectric circuits with elements like platinum (Pt), which are typically incompatible with these fragile crystals \cite{Shokri2024}.\\
Apart from realizing contacts, even the realization of heterostructures with ultra-thin materials is challenging due to issues like adsorbed molecules, which can lead to trapped bubbles or blisters between layers and cause misalignment and strain \cite{Cao2015}. This misalignment and strain can further degrade the electronic, optical, and mechanical properties of the heterostructure, reducing its effectiveness \cite{Gastaldo2023}. To address this challenge, one direction involves performing exfoliation in ultra-high vacuum (UHV) setups to minimize contamination and damage. Recent advancements include 
emerging exfoliation methods under UHV conditions to produce large-area monolayers \cite{ Sun2022, Grubisic2023, Haider2024}. \\
The innovations that are at the heart of the cuprate twistronics, both fundamental and applied, will likely lead not only to novel concepts and applications for quantum hardware, but also and most importantly, to a new way of conceiving how to engineer strongly correlated states in artificially made interfaces. Notable theoretical proposals that go beyond cuprates include iron-based superconductors \cite{Eugenio2023} and an intriguing combination of cuprates and iron-based superconductors for a 
multiband physics on demand \cite{Ricci2010}.\\

\noindent\textbf{Acknowledgements.} The authors thank S. Y. Frank Zhao, Philip Kim, Pavel Volkov, Jedediah H. Pixley and Marcel Franz for fruitful discussions. N.P. acknowledges the partial funding by the European Union (ERC-CoG, 3DCuT, 101124606), by the Deutsche Forschungsgemeinschaft (DFG 512734967, DFG 492704387, DFG 539383397 DFG 460444718, and DFG 452128813) and Terra Quantum AG. U.V. acknowledges funding by the European Union (ERC-StG, cQEDscope, 101075962), and by the Deutsche Forschungsgemeinschaft (DFG 539383397). V.B. and A.C. acknowledge support from Project PNRR MUR PE\_0000023-NQSTI financed by the European Union Next Generation EU.\\

\noindent\textbf{Author contributions.} N.P. and T.C. conceived and wrote the review with the main contributions of F.L.S and G. H. All the authors have discussed the manuscript.\\

\noindent\textbf{Declaration of Interest} All authors declare no conflict of interest.

\bibliographystyle{ieeetr}
\bibliography{bibliography}

\end{document}